\documentclass{article}
\usepackage{amsfonts, amsmath, amsthm, amssymb}
\usepackage[all]{xy}

\newcommand{\onesl}{\mathfrak{sl}_2}
\newcommand{\slsl}{(\mathfrak{sl}_2)_L\oplus(\mathfrak{sl}_2)_R}
\newcommand{\jl}{j_L}
\newcommand{\jr}{j_R}
\newcommand{\mb}{\mathcal{M}_\beta}
\newcommand{\mbhat}{\widehat{\mathcal{M}}_\beta}
\newcommand{\C}{\mathbb{C}}

\newcommand{\md}{\mathcal{M}_d}
\newcommand{\mdhat}{\widehat{\mathcal{M}}_d}
\newcommand{\mhat}{\widehat{\mathcal{M}}}
\newcommand{\sss}{\mathcal{E}}

\newcommand{\ih}{IH^\bullet(X)}
\newcommand{\HH}{\mathcal{H}}

\theoremstyle{remark}
\newtheorem*{remark}{Example}

\begin{document}

\title{Some remarks on Gopakumar-Vafa invariants.}
\author{A. Schwarz\thanks{Partly supported by NSF grant No. DMS 0204927.}, I. Shapiro}

\date{}
\maketitle

\begin{abstract}
We show that Gopakumar-Vafa (GV) invariants can be expressed in
terms of the cohomology ring of moduli space of D-branes without
reference to the $\slsl$ action.  We also give a simple
construction of this action.
\end{abstract}

\noindent\textbf{0.  Introduction}

\noindent Our goal is to express the Gopakumar-Vafa (GV)
invariants \cite{gv} of a three-dimensional Calabi-Yau manifold in
terms of the cohomology ring of moduli space of D-branes.  More
precisely, we consider the moduli space $\mb$ of holomorphic
curves in a Calabi-Yau 3-fold $M$ that belong to the homology
class $\beta\in H_2(M)$, the moduli space $\mbhat$ of
corresponding D-branes and the natural map
$p:\mbhat\rightarrow\mb$.  (To specify a D-brane wrapping a
holomorphic curve we should fix some additional data: a
holomorphic line bundle or, more generally, a semi-stable coherent
sheaf over the curve.)

One can construct an action of the Lie algebra $\slsl$ on the
$L^2$-cohomology $H^\bullet(\mbhat)$;  GV-invariants are defined
in terms of this action by the formula (\ref{gvdef}).

Consider an operator $L$ on $H^\bullet(\mbhat)$ acting as
multiplication by $p^*\omega$ where $\omega$ denotes the Kaehler
class of $\mb$.  We will show that one can obtain an expression
for GV-invariants in terms of $L$.  Define $\nu^\alpha_l$ as the
number of Jordan cells in the decomposition of $L$ having size $l$
and minimal degree $\alpha$.

We observe that  the character of the representation of $\slsl$ on
$H^\bullet(\mbhat)$ is recovered from $\nu^\alpha_l$ as
\begin{equation}
\chi(\varphi,\psi)=\sum_{l,\alpha} \nu^\alpha_l
e^{i(\alpha+l-1-d)\varphi}\dfrac{\sin(l\psi)}{\sin\psi}
\end{equation}
where $d=\text{dim}_\C\mbhat$.

To find GV-invariants $n_r$ one should represent the character in
the form
\begin{equation}\label{idecomp}
\chi(\varphi,\psi)=\sum a_{rs}4^{r+s} \cos^{2r}(\varphi/2)
\cos^{2s}(\psi/2)
\end{equation}
then $n_r=a_{r0}$.  Alternatively one can use the following
explicit formula
\begin{equation}
n_r=\sum_{\genfrac{}{}{0pt}{}{l,\alpha}{\alpha+l\geq
1+d}}(-1)^{l+1}l \nu^\alpha_l
(c^{\alpha+l-1-d}_r-c_r^{\alpha+l-3-d})
\end{equation}
where $c^j_r=(-1)^{r+j}{r+j+1\choose j-r}$ (see (\ref{decomp})).

The above statements can be derived from the considerations of
Sec.1 and from the existence of $\slsl$ action on the cohomology
of $\mbhat$.  The existence of such an action follows from
identification of this cohomology with the space of quantum BPS
states and the interpretation of BPS states in the framework of
M-theory.

It was conjectured in \cite{gv} that one can avoid any reference
to M-theory and construct this action by means of a Leray spectral
sequence associated to the map $p:\mbhat\rightarrow\mb$.  As was
shown in \cite{hst} the classical Leray spectral sequence should
be replaced by its perverse analogue (see \cite{bbd}) to give the
construction of the $\slsl$ action.

In Sec.2 we discuss the construction of $\slsl$ action using the
recent paper \cite{cm} that allows one to give a more detailed and
transparent picture than \cite{bbd}. We avoid using complicated
mathematical notions like perverse sheaves.  (We use the term
``intersection cohomology", but if one believes in the so called
Cheeger-Goresky-MacPherson conjecture, one can interpret
intersection cohomology as $L^2$-cohomology up to a shift.)

In Sec.3 we give a construction of the $\slsl$ action using
elementary linear algebra.  The hard theorems of \cite{bbd,cm} can
then be used to show that its character is identical to the one in
\cite{hst}.

\noindent\textbf{1.  Algebra of GV-invariants}

\noindent Let us consider representations of the direct sum of two
copies of the Lie algebra $\onesl$.  Irreducible representations
of $\slsl$ are labelled by two non-negative integers $\jl,\jr$. To
each such pair corresponds the tensor product $V_{\jl}\otimes
V_{\jr}$ where $V_{\jl}$ and $V_{\jr}$ denote the irreducible
representations of the left and right copy of $\onesl$ in $\slsl$.

The number $j$ stands for the highest weight of the
($j+1$)-dimensional representation $V_j$ of $\onesl$; physicists
use the spin $s=j/2$ to label representations.  The generators of
$(\onesl)_L$ and $(\onesl)_R$ will be denoted by $e_L, f_L, h_L$
and $e_R, f_R, h_R$ respectively; they satisfy the relations
$[e,f]=h, [h,e]=2e, [h,f]=-2f$.  Note that physicists would
re-scale our $h$ by $1/2$.

Denote by $I$ the $\onesl$ representation $V_1\oplus V_0\oplus
V_0$.  Write $I_r$ for the $r$-th tensor power of this
representation: $I_r=I^{\otimes r}$.  One defines GV-numbers $n_r$
of the $\slsl$ representation $V$ in the following way.  We
decompose $V$ into a direct sum:
\begin{equation}
V=\sum_r I_r\otimes R_r
\end{equation}
where $I_r$ is the representation of $(\onesl)_L$ defined above,
and $R_r$ is a (virtual) representation of $(\onesl)_R$.  Then
$n_r$ is defined by the formula
\begin{equation}\label{gvdef}
n_r=Tr_{R_r}(-1)^{h_R}
\end{equation}
If $V$ is decomposed into a direct sum of irreducible
representations:
\begin{equation}
V=\bigoplus_{\jl,\jr}N_{\jl,\jr}V_{\jl}\otimes V_{\jr}
\end{equation}
one can write down the following simple formula for the
GV-numbers:
\begin{equation}\label{nrex}
n_r=\sum_{\jl,\jr}(-1)^{r+\jl} {r+\jl+1\choose \jl-r}
(-1)^{\jr}(\jr+1)N_{\jl,\jr}
\end{equation}
This follows directly from the formula
\begin{equation}\label{decomp}
V_j=\sum_r (-1)^{r+j}{r+j+1\choose j-r}I_r
\end{equation}
where the RHS is interpreted as a direct sum of virtual
representations. (In other words (\ref{decomp}) should be
understood as an equality at the level of characters.)

To check (\ref{decomp}) we write down the characters:
\begin{equation}
\text{char}V_j=\dfrac{\sin((j+1)\phi)}{\sin(\phi)}\quad\text{and}\quad
\text{char}I_r=(2\cos(\phi/2))^{2r},
\end{equation}
then use the following well known formula ($n$ even)
\begin{equation}
\dfrac{\sin(na)}{\sin(a)}=\sum_{i=0}^\infty (-1)^i{n-i-1\choose
i}2^{n-2i-1}\cos^{n-2i-1}(a).
\end{equation}

Let us consider a finite dimensional graded vector space $V=\oplus
V^n$ equipped with an operator $L$ obeying $LV^n\subset V^{n+2}$.
Such a structure can be specified on the space of an $\slsl$
representation $V$ by taking the grading corresponding to the
diagonal Cartan operator $h_L+h_R$ and setting $L=e_R$.  We notice
that the character of the representation $V$ can be expressed in
terms of this structure.

Namely, observe that $V$ can be decomposed into a direct sum of
homogeneous cyclic $L$-modules, i.e. into a direct sum of
subspaces spanned by vectors $v, Lv, \cdots , L^{l-1} v$ where $v$
is homogeneous of degree $\alpha$.  We call such a subspace a
Jordan cell of size $l$ and minimal degree $\alpha$.  Let
$\nu^\alpha_l$ be the number of Jordan cells in this decomposition
having size $l$ and minimal degree $\alpha$.  Then the character
of $V$ is recovered by the formula
\begin{equation}
\text{char}(V)=\sum_{l,\alpha} \nu^\alpha_l
e^{i(\alpha+l-1)\varphi}\dfrac{\sin(l\psi)}{\sin\psi}.
\end{equation}

To check the above it is sufficient to consider $V=V_n\otimes
V_m$, it has $n+1$ Jordan cells of length $m+1$, with minimal
degrees $-n-m, -n-m+2, \cdots , n-m$.  One readily verifies the
formula in this case.

Corresponding GV-invariants can be calculated by using formula
(\ref{decomp}) and the observation that
$e^{-in\varphi}+e^{in\varphi}=\text{char}(V_n)-\text{char}(V_{n-2})$
for $n>0$. Explicitly
\begin{equation}
n_r=\sum_{\genfrac{}{}{0pt}{}{l,\alpha}{\alpha+l\geq
1}}(-1)^{\alpha+r}l \nu^\alpha_l \left[{\alpha+l+r\choose
2r+1}-{\alpha+l+r-2\choose 2r+1}\right].
\end{equation}
It is essentially the same formula as in the introduction but
without the shift by $d$ caused by the inconvenience of working
with the traditionally graded cohomology.

Now we make an observation to which we will return at the end of
Sec.2. Suppose that the character of an $\slsl$ representation is
given by $\chi(\varphi,\psi)=\sum h^{p,q}e^{ip\varphi}e^{iq\psi}$,
where $h^{p,q}$ are some integers.  It is easy to see that the
GV-invariants are then given by the formula:
\begin{equation}\label{hpqformula}
n_r=(-1)^r\sum_{p\geq r}(-1)^p\left[{p+r+1\choose
2r+1}-{p+r-1\choose 2r+1}\right]e(h^{p,\bullet})
\end{equation}
where $e(h^{p,\bullet})=\sum_q (-1)^q h^{p,q}$ is the Euler
characteristic.

For practical purposes it is inconvenient and unnecessary to find
a homogeneous Jordan decomposition of $V$ as described above.
Instead one can obtain all pairs $(l,\alpha)$ appearing in the
decomposition of $V$ via the following procedure.  Pick a vector
$v\in V$ of minimal degree, and let $V_v$ be the $L$ submodule
generated by $v$.  Read off $(l,\alpha)$ as the dimension of $V_v$
and the degree of $v$.  Consider $V/V_v$; again a graded
$L$-module.  Repeat until there is nothing left.

\begin{remark}  Let us consider a graded unital associative algebra $W$
generated by elements $x$ and $y$ of degree $2$ obeying the
relations $x^9=x^8 y+x^7 y^2$ and $y^3=0$. We define $L$ as the
multiplication by $x$.  Here we have to modify the discussion
above to introduce the $d$ back into it.

Taking $1\in W^0$ we get a pair $(10,0)$, taking $y\in W^2$ we get
$(10,2)$, finally $y^2\in W^4$ gives $(7,4)$.  From this we see
that $n_0=27$, $n_1=-10$ and the rest are $0$.  It is worthwhile
to observe that we are in the situation in which we can extend the
action of $L$ to that of $\slsl$ with the class of the
representation being $V_1\otimes V_9\oplus V_0\oplus V_6$.  It is
easiest to obtain the GV-invariants in this way whenever possible.

This example corresponds to the case of GV-invariants of the
manifold $M$ represented as the total space of $\mathcal{O}(-3)$
over $\C P^2$ \cite{example}. The algebra $W$ above is isomorphic
to the cohomology algebra of the space $\mbhat$ where $\beta=3\xi$
and $\xi$ is the generator of $H_2(M)=H_2(\C P^2)=\mathbb{Z}$.
The element $x$ comes from the Kaehler class $\omega$ of $\mb=\C
P^9$, i.e. $x=p^*\omega$.
\end{remark}

\noindent\textbf{2.  Geometry of GV-invariants}

\noindent GV-invariants of a Calabi-Yau threefold $M$ can be
defined in terms of the map $p:\widehat{\mathcal{M}}\rightarrow
\mathcal{M}$ where $\mathcal{M}$ stands for the moduli space of
holomorphic curves in $M$ and $\widehat{\mathcal{M}}$ denotes the
moduli space of corresponding D-branes. (To specify a D-brane one
needs a holomorphic curve and a (semi-stable) holomorphic vector
bundle or, more generally, coherent sheaf over the curve.) It
follows from physical considerations based on $M$-theory
interpretation of $\mhat$ that $\slsl$ acts on the cohomology of
$\mhat$. (This action can be obtained by identifying the
cohomology of $\mhat$ with the space of quantum BPS states.) Using
the action one can define GV-invariants.  More precisely, every
component of $\mathcal{M}$ specifies a sequence of GV-invariants.
The components $\mb$ of $\mathcal{M}$ are labelled by the homology
class $\beta$ of the holomorphic curve; one can identify $\mb$
with the corresponding component of the Chow variety.

It seems that the appropriate cohomology theory is
$L^2$-cohomology; this statement is supported by the fact that in
other situations it was successfully used to describe BPS states,
see for example \cite{ss}.  Cheeger, Goresky and MacPherson
\cite{cgm} conjectured that for a projective variety
$L^2$-cohomology coincides with intersection cohomology (more
precisely, with intersection cohomology with the middle
perversity).  In this conjecture $L^2$-cohomology is defined by
means of the standard metric on the projective space (Fubini-Study
metric) restricted to the smooth part of the variety.  If we
accept this conjecture, it is natural to define GV-invariants in
terms of the $\slsl$ action on the intersection cohomology of
$\mhat$.

One should observe that it is not quite clear how to define
rigorously the spaces $\mathcal{M}$ and $\widehat{\mathcal{M}}$
appearing in the definition of GV-invariants. (Physical
considerations determine only the part of these spaces
corresponding to non-singular curves.) It was suggested in
\cite{hst} to introduce $\mdhat$ as the normalized moduli space of
semi-stable sheaves $\sss$ of pure dimension $1$ on $M$ with
Hilbert polynomial $P(\sss,m)=d\cdot m+1$.  It was pointed out to
us by Sheldon Katz that using the cohomology of this moduli space
does not always lead to correct GV-invariants of a Calabi-Yau
3-fold.  There exists a natural map $\mdhat\rightarrow\md$ sending
every sheaf $\sss$ to its support, a curve of degree $d$, more
precisely to an element of the Chow variety. (It is known
\cite{hl} that $\mdhat$ is a projective scheme.) It was shown in
\cite{hst} that the $\slsl$ action on the intersection cohomology
of $\mdhat$ (more accurately, on the associated graded space of a
certain filtration on $IH^\bullet(\mdhat)$) can be obtained from
the Beilinson-Bernstein-Deligne theory of perverse sheaves
\cite{bbd}.

More generally, an action of $\slsl$ can be defined on the
intersection cohomology $\ih$ for any projective morphism
$f:X\rightarrow Y$ and ample line bundles over $X$ and $Y$. We
will discuss in more detail the $\slsl$ action on intersection
cohomology $\ih$ that corresponds to a projective map
$f:X\rightarrow Y$ of projective varieties and ample line bundles
$A$ over $Y$ and $\eta$ over $X$.  We use \cite{cm} that contains
a new proof of results in \cite{bbd} as well as some additional
useful facts.

It will be convenient for us to use the convention for the grading
of $\ih$ that places it in degrees between $-d$ and $d$, where
$d=\text{dim}_\C X$.  In this way the grading coincides with the
weights of the Cartan operator.

The line bundles $A$ and $\eta$ determine operators $L$ and $\eta$
on the cohomology $\ih$ defined by means of multiplication by
$f^*(c(A))$ and $c(\eta)$ respectively where $c(A)$ and $c(\eta)$
denote the characteristic classes of $A$ and $\eta$.  Note the use
of $\eta$ for both the line bundle and the associated operator.

It is well known that the operator $\eta^i:IH^{-i}(X)\rightarrow
IH^i(X)$ is an isomorphism.  For non-singular $X$ this fact (known
as the hard Lefschetz theorem) follows from standard theorems
about Kaehler manifolds.  Using this it is easy to construct
$\onesl$ action on $\ih$.  One starts by finding a homogeneous
basis $\{v_\alpha\}$ for the subspace of $\ih$ consisting of
primitive elements. ($v\in IH^{-i}(X)$  is primitive if
$\eta^{i+1}v=0$.) Let $\text{deg}v_\alpha=-i_\alpha$, then $\ih$
is a direct sum of subspaces spanned by $v_\alpha,\eta
v_\alpha,\cdots,\eta^{i_\alpha}v_\alpha$.  The representation of
$\onesl$ on $\ih$ is then defined by letting $e\in\onesl$ act as
$\eta$, the Cartan element $h\in\onesl$ act as multiplication by
the degree and the action of $f\in\onesl$ is defined inductively
by requiring that $fv_\alpha=0$.  In that way the direct sum
decomposition above becomes the decomposition of $\ih$ into
irreducible $\onesl$ submodules.

The operator $L$ (as does every nilpotent operator) specifies a
weight filtration $W_k$ on $\ih$.  Here it will be convenient to
diverge from the convention in \cite{cm} and use a decreasing
filtration defined as
\begin{equation}
W_k=\sum_{i+j=k}\text{Ker}L^{1-i}\cap\text{Im}L^{j}.
\end{equation}
It is characterized as the unique filtration with the properties
that $LW_i\subset W_{i+2}$ and $L^i:Gr_{-i}\rightarrow Gr_i$ is an
isomorphism.

Denote by $Gr^L\ih$ the associated graded space of the weight
filtration. Since $L:\ih\rightarrow IH^{\bullet+2}(X)$, the
subspaces $W_k$ of $\ih$ are homogeneous, i.e. $W_k=\bigoplus_i
W_k\cap IH^i(X)$, and so $Gr^L\ih=\bigoplus Gr^L_i IH^j(X)$; let
us set $\HH^{p,q}:=Gr^L_q IH^{p+q}(X)=W_q\cap
IH^{p+q}(X)/W_{q+1}\cap IH^{p+q}(X)$.

The operators $L$ and $\eta$ descend to $Gr^L\ih$.  We observe
that $L:\HH^{p,q}\rightarrow\HH^{p,q+2}$ and so
\begin{equation}\label{liso}L^i:\HH^{p,-i}\rightarrow\HH^{p,i}\end{equation}
is an isomorphism. (This
follows directly from the characterizing properties of the weight
filtration and the fact that $L$ preserves the $p$-degree.)  The
operator $\eta$ maps $\HH^{p,q}$ into $\HH^{p+2,q}$, and
\begin{equation}\label{etaiso}\eta^i:\HH^{-i,q}\rightarrow\HH^{i,q}\end{equation} is an
isomorphism.  This is a consequence of the identification of
$\HH^{p,q}$ with (following the notation of \cite{cm}) the so
called graded perverse cohomology groups $H^{d+p+q}_{-p}$, where
$d=\text{dim}_\C X$. This identification (in other words,
coincidence of the weight filtration of $L$ with the perverse
filtration associated with the map $f$) is one of the main results
of \cite{cm}. Similarly one can say that $\HH^{p,q}$ is identified
with the $E^{qp}_2$ term of the perverse spectral sequence
associated to the map $f$. (It is the same as the $ E^{qp}_\infty$
term.)  Now the isomorphism (\ref{etaiso}) is the so called
relative hard Lefschetz theorem of \cite{bbd,cm}.

Following an identical procedure to the one used in the
construction of the $\onesl$ action on $\ih$, we can use
isomorphisms (\ref{liso}), (\ref{etaiso}) to obtain an $\slsl$
action on $Gr^L\ih$.  In particular $e_L$ and $e_R$ act as $\eta$
and $L$ respectively and the Cartan operators $h_L$ and $h_R$
multiply $\HH^{p,q}$ by $p$ and $q$.  If $v\in\HH^{-p,-q}$ is a
primitive element (i.e. $L^{q+1}v=\eta^{p+1}v=0$) then the
elements $L^i\eta^j v$, $0\leq i\leq q$, $0\leq j\leq p$ span an
irreducible $\slsl$-invariant subspace of $Gr^L\ih$.  The lifting
of the $\slsl$ action to $\ih$ will be discussed later.

\begin{remark} Let us consider the case of $f:\widehat{M}_X\rightarrow M$,
where $M$ is a smooth projective variety, $X$ a smooth closed
subvariety, and $\widehat{M}_X$ is the blowup of $M$ along $X$.
Denote by $d$ the codimension of $X$ in $M$.  As usual, let $L$
denote the operator of multiplication by the pullback along $f$ of
the Chern class of an ample line bundle on $M$.  Then as an
$L$-module $H^\bullet(\widehat{M}_X)=H^\bullet(M)\oplus
H^\bullet(X)[-2] \oplus \cdots \oplus H^\bullet(X)[-2(d-1)]$,
where $L$ acts on the summands in an obvious way.

Thus it follows from the preceding discussion that the
GV-invariants are
$n_0=(-1)^{\text{dim}(\widehat{M}_X)}e(\widehat{M}_X)=(-1)^{\text{dim}(M)}(e(M)+(d-1)e(X))$
and for $r>0$, $n_r=(-1)^{\text{dim}(M)+r}{d+r-1\choose
2r+1}e(X)$.  The above formulas are most conveniently obtained by
observing that as an $\slsl$-module $Gr^L
H^\bullet(\widehat{M}_X)=V_0\otimes H^\bullet(M)\oplus
V_{d-2}\otimes H^\bullet(X)$.
\end{remark}

To reconcile the weight filtration approach with the discussion of
homogeneous Jordan cells in Sec.1 we observe that we may define
the weight filtration associated to the operator $L$ acting on $V$
as follows.  We decompose $V$ into a direct sum of cyclic
$L$-modules spanned by $v, Lv, \cdots , L^{l}v$ and define a
grading on $V$ by placing $v$ in degree $-l$ and setting the
degree of $L$ to be $2$.  This grading is admittedly non-canonical
however the associated filtration $W_k=\bigoplus_{i\geq k}V^i$
satisfies the characterizing property of the weight filtration of
$L$ and so is equal to it. Furthermore this grading gives an
$L$-equivariant isomorphism $V\rightarrow Gr^L V$ of graded
spaces.

This is readily modified to the case when $V$ was already graded
and $L$ had degree $2$ with respect to this grading by considering
the homogeneous Jordan cell decomposition.  Here we get a
bi-graded $V$ and an $L$-equivariant isomorphism
$V^{\bullet\bullet}\rightarrow Gr_\bullet^L V^\bullet$ of
bi-graded spaces.  As a consequence of the above we see that $\ih$
itself has a non-canonical bi-grading with total grading
coinciding with the usual one, and so there is an $L$-equivariant
isomorphism
$IH(X)^{\bullet\bullet}\rightarrow\HH^{\bullet\bullet}$ of
bi-graded spaces.  Unfortunately this isomorphism is non
$\eta$-equivariant, and so the $\slsl$ action is strictly speaking
constructed naturally only on $Gr^L\ih$, however it can be lifted
to $\ih$ via the above isomorphism.  In any case, if one is
willing to believe that  $\slsl$ acts on $\ih$ in such a way that
the grading corresponds to the action of the diagonal Cartan
operator $h_L+h_R$ and that $e_R$ acts by $L$, then the character
of this representation will be the same as the one obtained from
the action of $\slsl$ on $Gr^L\ih$ discussed here.

In light of the above discussion we see that the character of the
$\slsl$ action on $Gr^L\ih$ can be written down in terms of
$\HH^{p,q}$.  More precisely, let $h^{p,q}=\text{dim} \HH^{p,q}$,
then $\chi(\varphi,\psi)=\sum h^{p,q}e^{ip\varphi}e^{iq\psi}$, and
the GV-invariants are given by the formula (\ref{hpqformula}).  We
can interpret $e(h^{p,\bullet})$ as the Euler characteristic of
$^{perv}R^p f_* IC_X$, a perverse sheaf on $Y$.

\noindent\textbf{3.  A simple definition of the $\slsl$ action}

\noindent We begin by making the following general observation.
Let $A$ and $B$ be any two commuting nilpotent operators on a
vector space $V$.  Denote by $W_k$ the weight filtration of $V$
associated to $B$, then $AW_k\subset W_k$ and so $A$ and $B$
descend to $Gr^B V$, the associated graded space.  Furthermore,
let $W^i_k$ be the weight filtration of $Gr^B_i V$ associated to
$A$.  Recall that $B^i:Gr^B_{-i}V\rightarrow Gr^B_{i}V$ is an
isomorphism and now $B^i W^{-i}_k\subset W^i_k$ so that
\begin{equation}\label{trivisoA}
B^i:Gr^A_k Gr^B_{-i} V\rightarrow Gr^A_k Gr^B_{i} V
\end{equation} is an isomorphism.  For any $k$ we also have that
\begin{equation}\label{trivisoB}
A^i:Gr^A_{-i} Gr^B_{k} V\rightarrow Gr^A_i Gr^B_{k} V
\end{equation} is an isomorphism.  As a formal consequence of (\ref{trivisoA})
and (\ref{trivisoB}) (refer to Sec.2) we see that $Gr^A Gr^B V$
has a canonical structure of an $\slsl$-module with $e_L=A$,
$e_R=B$; $h_L$ and $h_R$ act on $Gr^A_p Gr^B_q V$ by $p$ and $q$
respectively.

In our particular situation let $V=IH(X)$ (the $\bullet$ is
missing to indicate that we forgot the grading), $A=\eta$ and
$B=L$.  We then get an $\slsl$ action on $Gr^\eta Gr^L IH(X)$ and
so may define GV-invariants of this action.  It is an immediate
consequence of the results in \cite{cm} that the character of the
representation above is the same as in \cite{hst}, and
consequently so are the GV-invariants.  (More precisely,
$\text{dim}Gr^\eta_p Gr^L_q
IH(X)=\text{dim}\HH^{p,q}=\text{dim}E^{qp}_2$, where $E^{qp}_2$
refers to the perverse spectral sequence associated to the map
$f$.) Note that if we reverse the order of the operators and
consider $Gr^L Gr^\eta\ih$, we do not get anything new, i.e. we
canonically obtain $\ih$ with the $(\onesl)_L$ acting trivially
and $(\onesl)_R$ acting via the hard Lefschetz.

\noindent\textbf{4.  Remarks}

\noindent \textbf{i)} In the construction of the $\slsl$ action we
have used operators $L$ and $\eta$, however the character of the
representation is determined by the bi-grading coming from the
perverse filtration on the $\ih$ associated to the projective map
$f:X\rightarrow Y$ and so \emph{depends only on the map}.
GV-invariants are defined in terms of the character of the
representation and so depend only on $f$ as well.

Let us consider a connected projective family of maps $f_s:
X_s\rightarrow Y_s$ of projective varieties labelled by a
parameter $s\in S$.  \emph{Then the corresponding GV-invariants
(and, more generally the character of the $\slsl$ representation)
do not depend on $s\in S$ if the cohomology of $X_s$ does not vary
over $S$}.  More precisely, given maps of projective varieties
$\mathcal{X}\xrightarrow{F}\mathcal{Y}\xrightarrow{\alpha} S$, for
every $s\in S$ we can consider maps $X_s\xrightarrow{f_s} Y_s$
where $Y_s=\alpha^{-1}(s)$, $X_s=F^{-1}(Y_s)$ and $f_s$ is the
restriction of $F$ to $X_s$. Let us assume that $S$ is connected
and for every $s\in S$ there is a neighborhood $U\subset S$
containing $s$ with the property that the inclusion of $X_s$ into
$X_U=(\alpha\circ F)^{-1}(U)$ induces an isomorphism of the
intersection cohomology.  Then the $\slsl$ representations
corresponding to the morphisms $f_s: X_s\rightarrow Y_s$ are
(non-canonically isomorphic) for all $s\in S$.
\vspace{\baselineskip}

\noindent \textbf{ii)} In the important paper \cite{kkv},
GV-invariants of a Calabi-Yau 3-fold $X$ were expressed in terms
of the Euler characteristics $e(C^{(\delta)})$, where
$C^{(\delta)}$ is roughly speaking the moduli space of holomorphic
curves of degree $d$ in $X$, together with a choice of $\delta$
points. (In \cite{kkv} it is assumed that these curves have
generic genus $g$.) It is interesting to note that our formula
(\ref{hpqformula}) is equivalent to Conjecture 3 (as it appears in
\cite{katz}) \emph{if we are able to verify} the equality
$$e(\mathcal{H}^{g-\delta,\bullet})=(-1)^{\text{dim} M +
\delta}(e(C^{(\delta)})-2e(C^{(\delta-1)})+e(C^{(\delta-2)})),$$
or its inverse
$$e(C^{(\delta)})=(-1)^{\delta+\text{dim} M}\sum_{i=0}^{\delta}(-1)^i (i+1)e(\mathcal{H}^{g-\delta+i,\bullet}),$$
where $e(\mathcal{H}^{p,\bullet})=\sum_q (-1)^q \text{dim}
\mathcal{H}^{p,q}$ (with $\mathcal{H}^{p,q}$ defined using the map
$f:\mdhat\rightarrow\md$) and can be interpreted as the Euler
characteristic of the perverse sheaf $^{perv}R^{p} f_*
IC_{\mdhat}$ on $\md$.

\vspace{\baselineskip} \noindent\textbf{Acknowledgements.} We are
indebted to J. Bernstein, D. Fuchs, S. Gukov, M. Kapranov, S. Katz
and A. Polishchuk for interesting discussions.

\noindent Department of Mathematics, University of California,
Davis, CA, USA \newline \emph{E-mail address}:
\textbf{schwarz@math.ucdavis.edu}
\newline \emph{E-mail address}:
\textbf{ishapiro@math.ucdavis.edu}
\end{document}